\documentclass[pra,twocolumn,showpacs,amsmath,amsfonts]{revtex4}
\usepackage{graphicx}

\newcommand{\bra}[1]{\langle #1 | \,}
\newcommand{\ket}[1]{\, | #1 \rangle}
\newcommand{\om}{\omega}
\newcommand{\Om}{\Omega}
\newcommand{\ga}{\gamma}

\newcommand{\ka}{\kappa}

\newcommand{\De}{\Delta}

\begin{document}

\title{Entanglement transfer from dissociated molecules to photons}

\author{David Petrosyan}
\author{Gershon Kurizki}
\author{Moshe Shapiro}
\affiliation{Department of Chemical Physics, 
Weizmann Institute of Science, Rehovot 76100, Israel}

\date{\today}

\begin{abstract}
We introduce and study the concept of a reversible transfer of the quantum 
state of two internally-translationally entangled fragments, formed by 
molecular dissociation, to a photon pair. The transfer is based on intracavity 
stimulated Raman adiabatic passage and it requires a combination of processes 
whose principles are well established. 
\end{abstract}


\maketitle

\section{Introduction}

The sharing of quantum information by distant partners in the form of 
their entanglement is the basis for quantum teleportation \cite{tlp}, 
cryptography \cite{crypt} and quantum computation \cite{comp}. The 
experimental and theoretical progress in entanglement generation and 
swapping \cite{swap}, has been impressive for two-state (spin-1/2-like) 
systems \cite{lloyd,demartini} or their translational (quadrature) degrees 
of freedom \cite{polzik}. Yet existing schemes for quantum teleportation 
are not suited to the formidable task of transferring quantum states of 
{\em complex material systems} such as molecules, to a distant node, where 
they can be recreated. Entanglement of spin or pseudospin states by 
dissociation has been studied for some time already \cite{spin}. Recently 
it has been suggested that dissociation into a translationally entangled 
pair of fragments, followed by a collision of one fragment with an atomic 
wavepacket, can be used to teleport the wavepacket \cite{tomas}. 

Here we put forward and study the concept of transferring the quantum state 
of two dissociated fragments sharing internal-translational entanglement 
to that of two photons and vice versa. Our proposal combines three schemes 
whose principles are well established: 
(a) the dissociation of a molecule into fragments, whose internal and 
translational states are naturally entangled (correlated)
\cite{phdiss,spin}; 
(b) the complete faithful mapping of the (unknown) states of the correlated 
fragments onto those of the photons, via intracavity stimulated Raman 
processes \cite{entgld,rempe}, performed in parallel in two cavities; 
(c) the transmission of the photons, tailored to form time-symmetric pulses, 
to two counterpart cavities at the distant node, where they produce the 
time-reversal of the aforementioned mapping, i.e., reproduce the quantum 
state of the fragments \cite{trnsfr}.

\section{State transfer from fragments to photons}

\begin{figure}[t]
\centerline{\includegraphics[width=8.5cm]{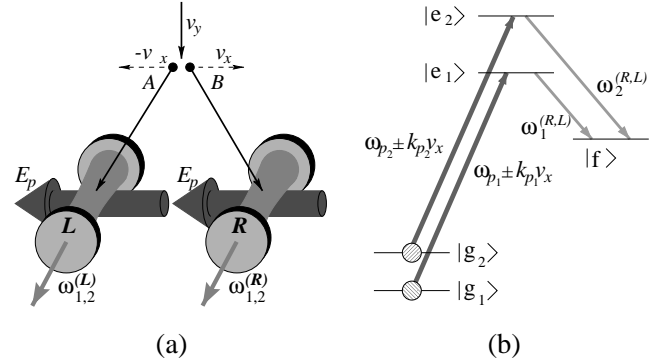}}
\caption{(a) Dissociating fragments $A$ and $B$ pass through the 
partially overlapping cavities $L$ and $R$ and pump fields, respectively, 
generating two correlated photons. The photons then leak from the cavities 
through the front mirrors. (b) Level scheme of dissociating fragments.}
\label{su_as1}
\end{figure}

The following procedure is envisaged for the task at hand (Fig. \ref{su_as1}). 
A cold molecule having velocity $v_y$ is dissociated, via a single- or 
two-photon process, to an energy-specific state of two identical molecular 
or atomic fragments, $A$ and $B$. Each fragment can occupy one of the two 
internal metastable states, labeled $\ket{g_1}$ or $\ket{g_2}$ 
(even if many internal states are populated by the dissociation process, we 
can single out the two that satisfy the resonance conditions detailed below). 
For a given dissociation energy, the fragments' velocities $\pm v_x$ along the 
$x$-axis depend on the internal excitation state of the system. Therefore we 
can place two empty optical cavities, $L$ and $R$, aligned along the $z$ axis, 
at positions such that only a pair of fragments in their single-excitation 
state $ \ket{g_1}_A \ket{g_2}_B  \pm \ket{g_2}_A \ket{g_1}_B$ enters both 
cavities, all other outcomes being idle events. This state is entangled and 
symmetrized or antisymmetrized, depending on the molecular configuration. 
Under the Raman-resonance condition, each fragment, passing through the 
sequence of partially overlapping cavity (quantized) and pump (classical) 
fields, undergoes population transfer to the final state $\ket{f}$, via 
stimulated Raman adiabatic passage (STIRAP) \cite{rempe}, and adds to the 
corresponding cavity a single photon at a frequency $\om^{(L,R)}_{1,2}$ 
{\it uniquely determined} by the initial internal and translational state. 
Upon leaving their cavities through the partially transparent front mirrors, 
the two entangled photons encode the dissociative state:
\begin{eqnarray}
& & (\ket{g_1,-p_x}_A \ket{g_2,p_x}_B \pm  
\ket{g_2,-p_x}_A \ket{g_1,p_x}_B ) \ket{0}_L \ket{0}_R 
\nonumber \\  & &
\to \ket{f,-p_x}_A \ket{f,p_x}_B  
(\ket{\om_1}_L \ket{\om_2}_R \pm 
\ket{\om_2}_L \ket{\om_1}_R ) , \label{ent}
\end{eqnarray}
where $\pm p_x$ are the momenta of the fragments in the center-of-mass frame. 

Let us now discuss a possible realization of the intracavity Raman 
resonance conditions [Fig. \ref{su_as1}(b)]. A bichromatic pump field, 
containing the frequency components $\om_{p_1}$ and $\om_{p_2}$ and aligned 
along the $x$-axis, resonantly couples the two ground states $\ket{g_1}$ and 
$\ket{g_2}$ with the excited states $\ket{e_1}$ and $\ket{e_2}$, respectively. 
Each cavity supports two modes with the frequencies 
$\om_{1,2}^{(L)} = \om_{p_{1,2}} - k_{p_{1,2}} v_x- (\om_f - \om_{g_{1,2}})$ 
and 
$\om_{1,2}^{(R)} = \om_{p_{1,2}} + k_{p_{1,2}} v_x- (\om_f - \om_{g_{1,2}})$,
where $\om_{p_{1,2}} = \om_{e_{1,2}} - \om_{g_{1,2}}$, 
$k_{p_{1,2}} = \om_{p_{1,2}}/c$ and 
$\hbar \om_j$ ($j=g_1,g_2,e_1,e_2,f$) are the energies of the corresponding
atomic states. Thus the mode frequencies of the two cavities are shifted 
from each other by the difference of the pump-field Doppler shifts for the 
two fragments, $\om_{1,2}^{(R)} - \om_{1,2}^{(L)} = 2 k_{p_{1,2}} v_x$, while 
the two modes of each cavity have a frequency difference close to that 
of the two excited levels, 
$\om_{2}^{(L,R)} - \om_{1}^{(L,R)} = \om_{e_2} - \om_{e_1} 
\mp (k_{p_2} - k_{p_1}) v_x \simeq \om_{e_2} - \om_{e_1}$. 
This choice ensures the two-photon Raman resonance between either of the 
states $\ket{g_1}$ or $\ket{g_2}$ and the final state $\ket{f}$ for both 
fragments.

The Hamiltonian of the system is given by
\begin{eqnarray}
H =& & \hbar  \sum_{j} \om_j \ket{j}\bra{j} 
+ \hbar \sum_{i} \{\om_i a_i^{\dag} a_i 
+ [\eta_i(t) \ket{e_i}\bra{f} a_i \nonumber \\ & &
- \Om_p (t) \ket{e_i}\bra{g_i} e^{-i (\om_{p_i} \mp k_{p_i} v_x) t} + 
\text{H.c.}]\},
\label{ham}
\end{eqnarray}
the upper (lower) sign in the exponent standing for fragment $A$ ($B$)
and cavity $L$ ($R$). Here the first term is the free-fragment Hamiltonian, 
where the sum is taken over all pertinent states, the second term
describes the cavity field, $a_i$ and $a_i^{\dag}$ ($i=1,2$) being the 
creation and annihilation operators for the corresponding mode, the third 
term describes the fragment-cavity interaction with the coupling $\eta_i(t)$ 
and the last term is responsible for the coupling of the fragment with the 
classical pump field, which is assumed to have the same Rabi frequency 
$\Om_p(t)$ on both frequency components $\om_{p_1}$ and $\om_{p_2}$. 

One of the requirements of STIRAP is the ``counterintuitive'' order of the 
fields \cite{bergmann}, achieved by shifting, by distance $d$, the pump-field 
maximum from that of the cavity field. For a fragment traveling with the 
velocity $v = \sqrt{v_x^2+v_y^2}$, the time-dependences of the cavity-(vacuum) 
and pump fields Rabi frequencies are then given by 
\begin{subequations}
\begin{eqnarray}
\eta_i(t) &=& \eta^{(i)}_0 \exp 
\left[- \left(\frac{v t}{w_c} \right)^2 \right] , \\
\Om_p(t) &=& \Om_0 \exp \left[- \left( \frac{v t - d}{w_p} \right)^2 \right] , 
\end{eqnarray}
\end{subequations}
where $\eta^{(i)}_0$ and 
$\Om_0$ are the corresponding peak Rabi frequencies and $w_c$ and $w_p$ are 
the waists of the cavity and pump fields. During the interaction, the 
combined system, consisting of the fragment plus its cavity field, will then, 
under the conditions specified below, adiabatically follow the ``dark'' 
eigenstate of the Hamiltonian (\ref{ham})
\begin{equation}
\ket{u^{(i)}_0(t)} = \frac{\eta_i(t) \ket{g_i,0} + \Om_p(t) \ket{f,\om_i} }
{\sqrt{\eta_i^2(t) + \Om_p^2(t)}} , \label{dark}
\end{equation}
which does not contain a contribution from the excited state $\ket{e_i}$ of 
the fragment. Thus, the fragment, being initially in state $\ket{g_i}$, ends 
up after the interaction in state $\ket{f}$, with a photon added into the 
corresponding cavity mode $\om_i$. 

A standard analysis \cite{bergmann} reveals the following requirements for 
the system to obey the evolution of the dark state (\ref{dark}): 
(i) The condition for adiabatic following should be satisfied, namely,
$\eta^{(i)}_0 w_c/v, \Om_0 w_p/v \gg \sqrt{1 + |\De_{p_i}|w_{c,p}/v}$, where 
$|\De_{p_i}| = k_{p_i} v_x$.
(ii) There has to be sufficient overlap between the two pulses:
$\int \eta_i(t) \Om_p(t) dt \gg \sqrt{\ga_{e_i}^2 + \De_{p_i}^2}$, where 
$\ga_{e_i}$ is the decay rate of the excited state $\ket{e_i}$. 
(iii) The fragment-cavity coupling strength should exceed the total 
relaxation rate of the combined final state $\ket{f,\om_i}$:
$|\eta^{(i)}_0| \gg \ga_f + 2\ka$, where $\ga_f$ is the decay rate of the 
fragment state $\ket{f}$ and $\ka$ is the transmission rate of the cavity 
field through the mirror, which is assumed to be the same for both cavity 
modes $\om_1$ and $\om_2$. 
(iv) Finally, the mode spacing of the cavity $|\om_2-\om_1|$ should exceed 
both the fragment-cavity coupling strength and the decay rate of the 
final state: $|\om_2-\om_1| \gg |\eta^{(i)}_0|,(\ga_f + 2\ka)$. Then, for a 
given initial state of the fragment $\ket{g_i}$, $i=1$ or $2$, the photon will
be added only into the resonant mode of the cavity at frequency $\om_i$.

The cavity mode operators obey the Langevin equation of motion \cite{gardiner}
\begin{equation}
\partial_t a_i = \frac{i}{\hbar} [H, a_i] - \ka a_i -
\sqrt{2 \ka} a_{\textrm{in}} , \label{lang}
\end{equation}
where $a_{\textrm{in}}$ is a quantum noise operator describing the input field.
The output field of each cavity is related to the input and internal fields of 
that cavity by 
\cite{gardiner}
\begin{equation}
a_{\textrm{out}} = a_{\textrm{in}} + \sqrt{2 \ka} (a_1 + a_2). \label{aout}
\end{equation}
Hence, for a vacuum input to the cavities, the output field is determined 
by the intracavity field, whose evolution is given by Eq. (\ref{lang}). 
Thus, according to (\ref{aout}), the entangled (correlated) state of the two 
intracavity fields is mapped onto the state of the outgoing photon pair. 
One can reconstruct the initial entangled state of the two dissociating 
fragments by making homodyne measurements of the output field of the two 
cavities for many repetitions of the dissociation process and using the
method of quantum state tomography \cite{tomography}.

For numerical analysis of the intracavity process and the photon transmission
we employ the density matrix formalism. In this approach, the time evolution 
of the system is governed by the master equation 
\begin{equation}
\partial_t \rho = -\frac{i}{\hbar} [H,\rho] - {\cal L} \rho ,
\end{equation}
where ${\cal L} \rho$ describes the fragment and cavity field relaxation 
processes. In this equation, several additional states of the system, which 
are decoupled from the Hamiltonian (\ref{ham}), have to be taken into account, 
as they enter through the possible relaxation channels. These are: 
$\ket{f,0}$---the fragment is in state $\ket{f}$ and the cavity is
empty; $\ket{l,\om_i}$ and $\ket{l,0}$---the fragment is in a low-lying 
state $\ket{l}$ to which the state $\ket{f}$ decays and the cavity has either
one or no photon. If the system successfully completes the transfer, then, 
irrespective of its initial state, it ends up in state $\ket{l,0}$ after a 
time long compared to all relaxation times, which corresponds to the absence 
of memory in the system about its initial state. The information about the 
initial state is transferred to the photons emitted by the corresponding 
cavities, as per Eqs. (\ref{ent}) and (\ref{aout}). 

\begin{figure}[t]
\centerline{\includegraphics[width=8.5cm]{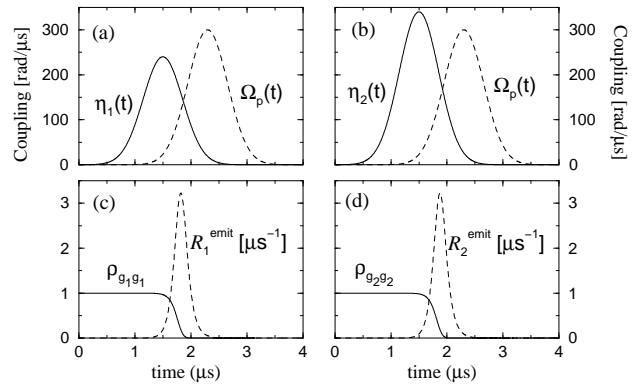}}
\caption{(a), (b): Time-dependence of the Rabi frequencies of the cavity 
$\eta_i(t)$ and pump $\Om_p(t)$ fields as seen by fragment $A$ being 
initially in state $\ket{g_1}$ (a) and fragment $B$ being initially in 
state $\ket{g_2}$ (b), or vice versa. 
(c), (d): Time evolution of the initial-state population 
$\rho_{g_1 g_1}$ (c) or $\rho_{g_2 g_2}$ (d) of the corresponding 
fragment and emission rate $R^{\text{emit}}_i$ of the photon from the cavity.
The parameters used are: $\ga_{e_{1,2}} \simeq 9.6$ MHz and 
$\ga_f \simeq 25$ MHz calculated for the transitions in text,
$\om_{p_1} / 2 \pi \simeq 9.0738 \times 10^{14}$ s$^{-1}$ 
and $\om_{p_2} / 2 \pi \simeq 9.0755 \times 10^{14}$ s$^{-1}$, 
$k_{p_{1,2}} v_x/2\pi \sim 15$ MHz, $(\om_2 - \om_1)/2\pi \simeq 168.9$ GHz, 
corresponding to the mode spacing of a cavity 0.9 mm long. The corresponding 
coupling constants for the cavity having the mode-waist $w_c \sim 10$ $\mu$m
are $\eta^{(1)}_0/2 \pi \simeq 38$ MHz and $\eta^{(2)}_0/2 \pi\simeq 54$ MHz 
and the cavity linewidth $2\ka \simeq 10$ MHz. Reduction of the 
fragment-cavity coupling constants by a factor of 4 (!) lowers the transfer 
efficiency from 99\% to 90\%.}
\label{graph}
\end{figure}

For numerical simulations of the system's dynamics we have chosen a 
well-collimated, cold beam of sodium dimers. Such a beam can be produced via 
stimulated Raman photoassociation of cold Na atoms, thereby obtaining 
translationally cold Na$_2$ molecules in the chosen vib-rotational state of 
the electronic ground state $X^1\Sigma^+_g$ \cite{bec_mol}. Subsequently, the 
inverse Raman process dissociates the molecules into pairs of 
internally-translationally entangled fragments sharing a single excitation 
\cite{vardi}. The cavities admit fragments with
$v_x \simeq 5$ m/s and $v_y \simeq 10$ m/s. The two frequency components of 
the pump field couple the two metastable ground states of the Na atom 
$\ket{g_1}=\ket{3S_{1/2},F=1}$ and $\ket{g_2}=\ket{3S_{1/2},F=2}$ with the 
excited states $\ket{e_1}=\ket{4P_{1/2}}$ and $\ket{e_2}=\ket{4P_{3/2}}$, 
respectively. The final state is $\ket{f}=\ket{4S_{1/2},F=2}$. 
The sequences of fields seen by the fragments in cavities with parameters
similar to those of Ref. \cite{rempe} are plotted in Figs. \ref{graph}(a) 
and \ref{graph}(b), while the populations of the initial states $\ket{g_i}$ 
and the corresponding photon emission rates, defined as 
$R^{\text{emit}}_i = 2 \ka (\rho_{f,\om_i;f,\om_i} + \rho_{l,\om_i;l,\om_i})$,
are plotted in Figs. \ref{graph}(c) and \ref{graph}(d). For both cavities, 
the total photon emission probability 
$P_i = \int R^{\text{emit}}_i dt \geq 0.99$, indicating extremely high 
efficiency (fidelity) of entanglement transfer between the fragments and the 
photons, as per Eq. (\ref{ent}).  It is noteworthy that this efficiency 
(or fidelity) remains very high even for considerably lower fragment-cavity
coupling strengths (Fig. \ref{graph} caption). 

After the fragments have interacted with the corresponding cavity and pump 
fields and the generated photons have propagated away from the cavities, the 
system is reset to its initial state. We can then repeat the process, 
generating a second pair of photons, and so on. The time interval between  
two subsequent photon-pair transmissions must exceed 
$T \simeq \max[(w_c+w_p)/v,(2\ka)^{-1}]$, which limits their maximum 
repetition rate. Since on average half of the dissociation events are idle, 
yielding a pair of fragments that are both in either $F=1$ or $F=2$ states, 
the actual repetition rate is given by $W \leq (2T)^{-1}$. With the parameters 
of Fig. \ref{graph}, the maximal repetition rate is $W \simeq 200$ kHz.
Since the fragments' velocities $\pm v_x$ along the $x$-axis depend on the 
internal excitation state of the system, one can resort to post-selection,
by detecting only those pairs of fragments that have successfully crossed the 
cavities and generated an entangled photon pair, thereby discarding all other
idle events. Such post-selection, together with the fact that the fidelity of 
entanglement transfer between the fragments and the photons is close to 1, 
would make our scheme deterministic, rather that probabilistic, as opposed to 
the spontaneous parametric down-conversion schemes \cite{spdc}. 

The outlined processes are also feasible for molecular dissociation into
two molecular fragments in the electronic ground state. As an example 
consider the photolysis of cyanogen by a 193 nm laser via the reaction channel
$\mbox{C}_2\mbox{N}_2(X^1\Sigma^+_g) + \hbar \om_{\text{diss}}\to 
\mbox{CN}(X^2\Sigma^+,v_1 = 0, N_1 \leq 45) +
\mbox{CN}(X^2\Sigma^+,v_2 = 1, N_2 \leq 31)$, where $v_{1,2}$ and $N_{1,2}$
stand for vibrational and rotational states, respectively, of the electronic 
ground state $X^2\Sigma^+$ of CN \cite{C2N2}. By adjusting the positions of 
the two cavities and the frequencies of the cavity and pump fields, the two 
ground states $\ket{g_{1,2}}$ and the final state $\ket{f}$ can be selected 
from this vib-rotational ground state manifold,  while the excited states 
$\ket{e_{1,2}}$ can be selected from among the excited electronic state 
manifold $B^2\Sigma^+$. 

A possible conceptual counterargument for the use of our scheme may be that, 
depending on the initial state $\ket{g_1}$ or $\ket{g_2}$, the fragment recoil 
due to the pump-photon absorption will be $\hbar k_{p_1}$ or $\hbar k_{p_2}$, 
respectively. A subsequent measurement of the fragment's translational state 
will, in principle, disclose its initial internal state. Consequently, the 
final motional states of the two fragments will be entangled with the states 
of the two generated photons, without achieving a complete state mapping from 
the fragments onto the cavity photons. However, one can easily check that, 
since the molecule is dissociated in a region having the size 
$D_x \leq w_c \simeq 10$ $\mu$m, in order to have each dissociating fragment 
pass through the corresponding cavity waist, the uncertainty of the momentum 
distribution of the fragment must satisfy $\De p_x \agt \hbar / D_x$, which 
is 30 times larger than the photon-recoil difference $\hbar (k_{p_2}-k_{p_1})$.
Therefore, even in principle one will not be able to resolve that difference 
and deduce the initial state from the fragments' momenta. 

We note that a fragment crossing a standing wave cavity at a node, where
the electric field amplitude vanishes, will not interact with the cavity 
mode and the STIRAP process will not take place. One possibility to 
overcome this difficulty is to allow the fragment to cross the cavity axis 
at an angle slightly different from 90$^\circ$, which can be achieved by 
tilting the cavity. Another possibility would be to use a running-wave
cavity.

\section{State transfer between distant nodes}

\begin{figure}[t]
\centerline{\includegraphics[width=8.5cm]{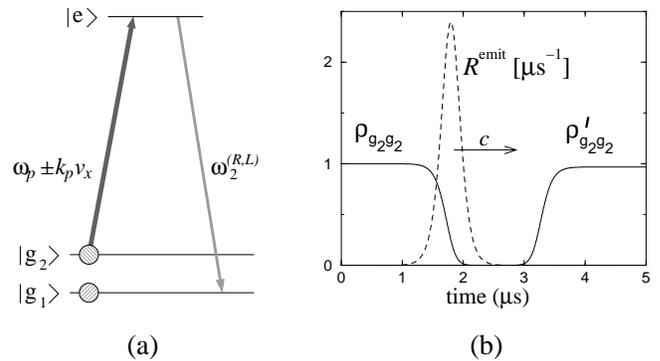}}
\caption{(a) Level scheme of dissociating fragment interacting with a single 
mode of the corresponding cavity at frequency $\om_2^{(L,R)}$. 
(b) Transfer of the population of state $\ket{g_2}$ between fragments 
$A$ ($B$) and $A^{\prime}$ ($B^{\prime}$) via a single photon.
The parameters used are: $\om_p / 2 \pi \simeq 5.083 \times 10^{14}$ s$^{-1}$, 
$k_p v_x/2\pi \sim 8.5$ MHz, $\eta_0/2 \pi \simeq 22$ MHz, and 
$\ga_e \simeq 6.28$ MHz. 
All other parameters are the same as in Fig. \ref{graph}.}
\label{as_gr}
\end{figure}

So far we have only considered the state transfer from a pair of dissociating 
fragments to photons. Utilizing the transmission protocol of 
Ref. \cite{trnsfr}, one may use the generated entangled photon pair to induce 
the {\it inverse process at a distant node}, so as to convert the dissociating 
state $\ket{f,-p_x}_{A^{\prime}} \ket{f,p_x}_{B^{\prime}}$ of another pair of 
fragments, $A^{\prime}$ and $B^{\prime}$, into the initial state of fragments 
$A$ and $B$. This procedure is applicable to molecular fragments in the 
electronic ground state, but not to atomic fragments whose final state 
lifetime $\ga_f^{-1}$ is shorter than their time of flight between the 
dissociation region and the corresponding cavity. Let us therefore consider 
an alternative {\it simplified scheme}. Each cavity in Fig. \ref{su_as1}(a) 
now supports only one mode at frequency 
$\om_2^{(L,R)} = \om_e - \om_{g_1} \mp k_p v_x$, respectively. Together with 
a monochromatic pump field having a frequency $\om_p = \om_e - \om_{g_2}$, 
this provides the two-photon Raman resonance for each fragment between the 
states $\ket{g_2}$ and $\ket{g_1}$ [Fig. \ref{as_gr}(a)]. Upon passing through 
the cavity and pump fields, only a fragment initially in state $\ket{g_2}$ 
will undergo the intracavity STIRAP to state $\ket{g_1}$ and add a photon to 
the corresponding cavity. Due to the large two-photon Raman detuning 
$\om_{g_2} - \om_{g_1} \gg \eta_0, \Om_0$, a fragment occupying initially 
state $\ket{g_1}$ will exit the interaction region in the {\it same state}. 
Thus, both fragments will end up in state $\ket{g_1}$. 
The fragment-cavity Hamiltonian takes the form
\begin{eqnarray}
H =& & \hbar  \sum_{j} \om_j \ket{j}\bra{j} 
+ \hbar \om_2 a^{\dag} a + \hbar [\eta(t) \ket{e}\bra{g_1} a \nonumber \\ & &
- \Om_p (t) \ket{e}\bra{g_2} e^{-i (\om_{p} \mp k_p v_x) t} + \text{H.c.} ].
\label{ham_tr}
\end{eqnarray}
Here we assume that the {\it momentum uncertainty} of the fragment is 
{\it large} and exceeds the photon momentum, $\De p_x > \hbar k_p$. We, 
therefore, neglect the recoil of the fragment due to the absorption of the 
pump photon, obtaining
\begin{eqnarray}
& & (\ket{g_1,-p_x}_A \ket{g_2,p_x}_B  
\pm  \ket{g_2,-p_x}_A \ket{g_1,p_x}_B ) \ket{0}_L \ket{0}_R 
\nonumber \\  & &
\to \ket{g_1,-p_x}_A \ket{g_1,p_x}_B  
(\ket{0}_L \ket{\om_2}_R \pm 
\ket{\om_2}_L \ket{0}_R ) . \label{ent2}
\end{eqnarray}

The generated photon leaks out of the corresponding cavity at the rate 
$2 \ka$. Let the output of the cavities $L$ and $R$ be directed through, say, 
an optical fiber into two similar cavities $L^{\prime}$ and $R^{\prime}$ 
constituting the receiving node of the system. At that node, a molecule having 
the same velocity $v_y$ is dissociated to produce two fragments $A^{\prime}$ 
and $B^{\prime}$ in the state $\ket{g_1}_{A^{\prime}} \ket{g_1}_{B^{\prime}}$. 
The dissociation energy of the $A^{\prime}-B^{\prime}$ molecule is reduced 
relative to that of the $A-B$ molecule, by an amount equal to the energy 
separation between the two ground states $\ket{g_1}$ and $\ket{g_2}$, so that 
the dissociating fragments $A^{\prime}$ and $B^{\prime}$ have the same 
velocities $\mp v_x$ as fragments $A$ and $B$. The two pulsed dissociation 
processes are appropriately synchronized, so that the fragments $A^{\prime}$ 
and $B^{\prime}$ pass through the cavities $L^{\prime}$ and $R^{\prime}$ when 
they receive the output of the cavities $L$ and $R$. Time-reversal is 
achieved by allowing the fragments $A^{\prime}$ and $B^{\prime}$ to interact 
first with the pump field and then with the cavity field. Provided the 
inversion process is successful, the fragments $A^{\prime}$ and $B^{\prime}$ 
will end up in the {\it same initial internal-translational state} as 
fragments $A$ and $B$ (before the interaction with their cavities).

We have studied the dynamics of the system composed of the two distant nodes 
using the density operator formalism developed in \cite{ucpl}. The master 
equation now reads 
\begin{equation}
\partial_t \rho = -i \hbar^{-1} [H + H^{\prime},\rho] - {\cal L} \rho -
2\ka ([a^{\prime \dag}, a \rho] + [\rho a^{\dag}, a^{\prime}] ) ,
\end{equation} 
where the primed operators stand for the receiving node, ${\cal L} \rho$ 
describes the fragment and cavity field relaxation processes at both nodes, 
and the last term provides the unidirectional coupling between the two nodes, 
in which the output of the cavity at the sending node constitutes the retarded 
input for the cavity at the receiving node. As expected \cite{trnsfr}, our 
simulations show that, provided the photon wavepacket $R^{\text{emit}}$ is 
completely time-symmetric, the processes at the two nodes are the time 
reversals of each other. Figure \ref{as_gr}(b) illustrates the results of our 
calculations. The parameters used again correspond to a dissociating sodium 
dimer and the excited state $\ket{e}$ corresponds to the state 
$\ket{3P_{1/2},F=2}$ of the Na atom. The probability of transferring the 
population of state $\ket{g_2}$ from fragment $A$ ($B$) to fragment 
$A^{\prime}$ ($B^{\prime}$) and thereby achieving the reversal of 
Eq. (\ref{ent2}) is calculated to be 97~\%. 
It is the decay of the excited atomic state $\ket{e}$ that reduces the 
fidelity of the process from 100~\% to 97~\%. Other sources of decoherence, 
such as photon absorption in the mirror and during propagation, can be 
accounted for by introducing an additional relaxation channel with a loss rate
$\ka^{\prime}$ \cite{trnsfr}. A simple analysis shows that the fidelity
of the process is proportional to $\ka/(\ka + \ka^{\prime})$, which is
also confirmed by our numerical simulations.

\section{conclusions}

In this paper we have proposed the hitherto unexplored possibility of probing
and exploiting the quantum information associated with internal-translational 
entanglement in molecular dissociation. Our scheme allows, in principle, 
high-fidelity state transfer from the entangled dissociated fragments to light,
thereby producing a highly correlated photon pair. This process can be 
followed by its reversal at a distant node of a quantum network resulting in 
the recreation of the original two-fragment entangled state. The proposed
process may have advantageous applications in quantum teleportation and
cryptography. Thus, the quantum information encoded in the entangled
dissociative state can be shared by two distant partners who will each
possess half of a crypto-key. This key will evidently be sensitive to 
tampering by eavesdropping owing to the high fidelity of its preparation. 
We would like to stress that the proposed scheme requires a combination of 
processes whose principles are well established: preparation of 
translationally cold dimers in a chosen vib-rotational level of the
electronic ground state \cite{bec_mol}, one- or two-photon dissociation
\cite{phdiss} of the dimer via an energy-specific potential surface 
whereby the identical fragments with selected velocities are in an
entangled state as per Eq. (\ref{ent}), and intracavity STIRAP for each 
fragment \cite{rempe,bergmann}.

\begin{acknowledgments}
We acknowledge the support of the EU QUACS RTN (G.K.), the Feinberg 
School (D.P.) and the EU COCOMO RTN (M.S.).
\end{acknowledgments}

\end{document}